\documentclass[aps,pra,reprint,twocolumn,superscriptaddress]{revtex4-2}

\usepackage{graphicx}
\usepackage{dcolumn}
\usepackage{bm}

\usepackage{amsmath}
\usepackage{amsfonts}
\usepackage{braket}
\usepackage{lineno}
\usepackage{xcolor}
\usepackage[section]{placeins}

\begin{document}

\title{Phase-controlled transport of Floquet-driven compact chiral photonic states}

\author{Gabriel Cáceres-Aravena$^\dagger$}\email{gabriel.caceres-aravena@uni-rostock.de}\affiliation{Institute of Physics, University of Rostock, 18051 Rostock, Germany.}
\author{Paloma Vildoso}\thanks{These two authors contributed equally to this work}\affiliation{Departamento de Física, Facultad de Ciencias Físicas y Matemáticas, Universidad de Chile, Santiago, Chile.\\Millennium Institute for Research in Optics--MIRO, Santiago, Chile.}
\author{Helena Dr\"ueke}\affiliation{Institute of Physics, University of Rostock, 18051 Rostock, Germany.}
\author{Rodrigo A. Vicencio}\affiliation{Departamento de Física, Facultad de Ciencias Físicas y Matemáticas, Universidad de Chile, Santiago, Chile.\\Millennium Institute for Research in Optics--MIRO, Santiago, Chile.}

\date{\today}

\begin{abstract} 
The Aharonov-Bohm (AB) effect remains a cornerstone of fundamental and applied physics. In this work, we utilize the AB caging effect originated from an effective magnetic field induced by multiorbital interactions, creating an all flat band (FB) lattice system. Normally, FB states are known for being compact in space and having a zero tail; therefore, their mobility in a linear environment is generally understood as impossible. We propose a Floquet driving protocol in an all-FB photonic system to fully control the dynamics of localized photonic states. The modulation of the Hamiltonian along the propagation coordinate allows the translation of compact states in the direction of constructive interference, resulting in an effective stroboscopic quantum walk-like effect. We find that the traveling states exist in chiral pairs. 
We experimentally implement the Floquet driven protocol using femtosecond laser written photonic waveguides and demonstrate directional control of the propagation, determined by the relative phase of the input condition. 
\end{abstract}

\maketitle 

\section{Introduction}

One of the main challenges in physics is the localization, transport, and manipulation of energy, as their controlled realization form the fundamental building blocks of functional materials and devices. 
The potential to achieve this with speed-of-light transmission has made photonics a field of growing interest, not only due to the expected processing speed, but also because it has been demonstrated as a versatile platform for signal processing, cryptography, among many others.
Moreover, photonic lattices offer on-chip wave propagation, enabling compact and scalable implementations of various models due to their flexibility to construct complex two-dimensional (2D) configurations~\cite{LEDERER20081}. In addition, photonic lattices have been extensively studied to explore different phenomenaphenomena such as flat bands (FBs)~\cite{Vicencio_FBdynamics,LeykamFlach_FB,chen2024chip,10.1063/5.0153770,vicencio_observation_2015,seba2015,SebaOL,Yang2024Feb,DanieliAndreanovLeykamFlach,doi:10.1142/S021797921330017X,PhysRevLett.121.263902}, Floquet engineered systems~\cite{maczewsky_observation_2017,PhysRevB.82.235114,rechtsman_photonic_2013}, photonic topological insulators~\cite{grafeno24,rechtsman_photonic_2013,refId0,doi:10.1126/science.abd2033}, photonic quantum simulators~\cite{grafeno24,doi:10.1126/science.abd2033,rechtsman_photonic_2013,PhysRevA.102.023505,mazanov_photonic_2024}, and integrated photonic gates~\cite{crespi_integrated_2011,RevModPhys.79.135,doi:10.1126/science.1155441}. A key element in photonic lattices is that their dynamical evolution can be accurately described by tight-binding-like Hamiltonians, making them a highly tunable platform to simulate condensed matter physics~\cite{https://doi.org/10.1002/lpor.200810055}.
In the context of energy localization, FB systems have emerged as key solutions due to their localization properties, 
as the energy can be perfectly trapped in a very precise and compact form~\cite{Vicencio_FBdynamics,LeykamFlach_FB}. 
FB states are linear solutions which have no tails due to the destructive interference of amplitudes at connector sites~\cite{FBluis}. Therefore, they are spatially compact and occupy a reduced number of lattice sites only, covering one or few unit cells.
In recent studies, the development of multiorbital interactions in photonics has allowed for the creation of controlled negative hoppings between optical guided modes in a simple and robust manner~\cite{guzman-silva_experimental_2021,CaceresAravena2022Jun,mazanov_photonic_2024,grafeno24,Multiorbitalphotoniclatt}. The addition of second-order modes could produce negative hoppings that favors FB generation due to destructive interference; therefore, multi orbital states appear to be intrinsically connected to FB systems.
The addition of this negative interorbital coupling on a lattice allows the creation of robust all-FB systems~\cite{Yang2024Feb}, which show complete localization for any input excitation~\cite{CaceresAravena2022Jun}. 
However, a promising technique to manipulate the dynamics relies on Floquet engineering~\cite{floquet_sur_1883,https://doi.org/10.1002/adpr.202400023,li_topological_2022,lindner_floquet_2011,della_valle_floquet-hubbard_2014,zhu_floquet-surface_2020,PhysRevB.91.235134,PhysRevB.73.073104,ahmed2025,PhysRevResearch.5.023056}. By subjecting the system to periodic driving, we can generate synthetic Hamiltonians with novel dynamical properties, which are otherwise inaccessible in static configurations. 
Moreover, the inclusion of a time dependency of the Hamiltonians can lead to topological properties~\cite{PhysRevB.82.235114,PhysRevLett.116.156803,Lopez_2020}. 

In this work, we propose a mechanism to manipulate bulk excitations in a controlled manner, without experiencing spatial dispersion of the wave packets. We use a diamond (also called rhombic) lattice with induced synthetic magnetic fluxes that produce an all-FB spectra~\cite{CaceresAravena2022Jun}. One of the main advantages of using this lattice and Aharonov-Bohm (AB) caging effects~\cite{AB59,PhysRevLett.81.5888} is that the system is also robust against static or time-periodic disorder~\cite{PhysRevA.101.023839}. 
The experimental demonstration is performed in femtosecond-written photonic lattices~\cite{Szameit2010}, incorporating interorbital interactions to induce a synthetic magnetic flux of $\pi$~\cite{CaceresAravena2022Jun,Multiorbitalphotoniclatt}. 
The primary advantage of femtosecond-written waveguides is the versatility of the lattices that can be built. This fabrication method has been used to study not only multiorbital interactions~\cite{CaceresAravena2022Jun,Multiorbitalphotoniclatt}, but also topological parity-time-symmetric non-Hermitian systems~\cite{weimann_topologically_2017}, anomalous Floquet topological insulators~\cite{maczewsky_observation_2017,koyama_designing_2023}, and recently this method was adopted for dense, fast and efficient archival data storage in glass~\cite{allison_laser_2026}. 
Our experimental results demonstrate the controlled transport of compact FB states by implementing a Floquet protocol based on the concatenation of two photonic masks. Although experimental observations show certain leakage mechanisms arising from fabrication asymmetries, the predominant signal remains clearly identifiable, allowing for the successful observation of the transport dynamics. Our results confirm that the Floquet-engineered system effectively drives the movement of FB states, where the directionality is governed by interference at the connector sites.

\section{Theoretical Description}

We study the evolution of light along the propagation coordinate $z$ in coupled waveguide arrays~\cite{Szameit2010} by analyzing the dynamics of the electric field envelopes at different lattice sites. We specifically consider single mode ($S$) and multi mode ($S$-$P$) waveguides, such that we can tune the interorbital coupling between the $S$ and the $P$ photonic modes at adjacent sites~\cite{guzman-silva_experimental_2021}. Depending on the waveguides orientation, this allows us to induce positive and negative coupling constants and, therefore, to effectively adjust the lattice dynamical properties. For instance, we can transit from a mostly dispersive system into a zero dispersion (all-FB) one. In other words, the inclusion of the interorbital coupling can be interpreted as the photonic implementation of an effective magnetic field in the lattice, and the induction of AB caging. 

\begin{figure}[!ht]\centering
\includegraphics[width=0.48\textwidth]{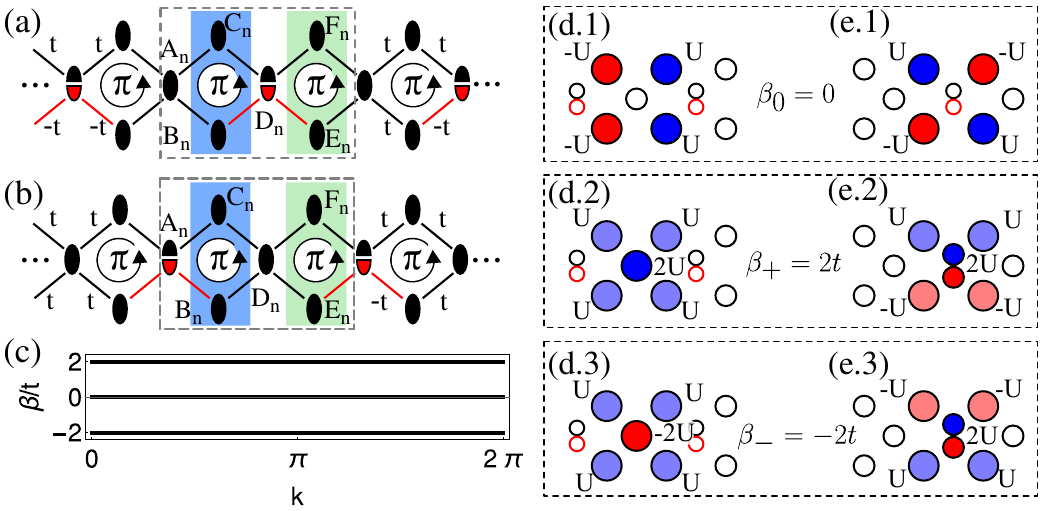}
	\caption{(a,b) The two configurations of the diamond  model. The gray rectangle highlights the unit cell. The waveguide types are: completely black for $S$ sites and half black half red for the $P$ sites. (c) Bloch Bands $\beta/t$. (d,e) Sketch of the FB states for the eigenvalues $\beta=0,\ 2t$, and $-2t$. Blue and red colors represent normalized positive and negative amplitudes, respectively. The $P$ modes are represented by half-blue-half-red colored solid circles at the center.}\label{fig1}
\end{figure}

We start by analyzing the static system using Bloch theory~\cite{Bloch1929}. The effect of the interorbital coupling on the dispersion properties of a diamond lattice, see Figs.~\ref{fig1}(a) and (b), can be fully understood by analyzing a static configuration.
In both cases, we have a diamond-like lattice with a synthetic magnetic field induced by an $SP$ interaction~\cite{CaceresAravena2022Jun}. The lattice is formed by two different sites: the $S$ one (black in the figures), which hosts the fundamental ($S$) photonic mode only; and the $P$ site (black and red in the figures), which supports the fundamental and the second order ($P$) modes. The $SP$ interaction is enhanced by finely tuning the propagation constants of the $S$ and the $P$ modes at different lattice sites, whereas the $S$ mode at the $P$ site is effectively decoupled from the rest of the system~\cite{Cantillano_2018}. The dashed rectangles in Figs.~\ref{fig1}(a,b) correspond to the lattice unit cell composed of six sites. Black lines illustrate the coupling constant ``$t$'', while the red lines describe the negative hoppings ``$-t$''. Depending on the orientation of the mode orbitals~\cite{Multiorbitalphotoniclatt}, the coupling among different sites can take a positive or negative value. The coupling among $S$ modes is always defined positive, while the one in between the $S$ and the $P$ states will strongly depend on the angle; for example, the coupling is exactly zero on an horizontal disposition.

The highlighted blue and green areas are of great relevance in this work: initial conditions in those regions will transform into traveling states, which is a crucial and counterintuitive result. The diamond homogeneous ($S$-only) lattices have a spectrum formed by one flat and two dispersive bands~\cite{SebaOL}, where the FB compact states consist of two out-of-phase amplitudes exactly located at the highlighted areas of Figs.~\ref{fig1}(a,b). 
In a homogeneous system, exciting these compact states leads to them remaining localized. This is because a perfect destructive interference mechanism occurs at the left and right connector (central $A,D$) sites. However, in our modulated $SP$ scheme this out-of-phase profile and, also, an in-phase configuration, will be both a key for successfully producing a controlled transport mechanism~\cite{CaceresAravena2022Jun}.

We consider a tight-binding (coupled-mode) approximation~\cite{LEDERER20081} to study the $SP$ diamond lattices described in Figs.~\ref{fig1}(a,b). For this purpose, we define a Bloch vector as
\begin{equation}
	\psi = ( \psi_A , \psi_B , \psi_C , \psi_D , \psi_E ,\psi_F )^T \ ,
\end{equation}
with $\psi_i$ representing the mode amplitude at the respective lattice sites. 
Every optical mode has a propagation constant $\beta_{S,P}$ along the $z$ coordinate, which is the dynamical variable. In this case, we tune the waveguides properties such that $\beta_S=\beta_P$ at different sites~\cite{guzman-silva_experimental_2021}. The propagation constant is represented by diagonal terms in the Hamiltonian. These terms are eliminated from the matrix using a trivial phase transformation.
The equations governing the dynamics of this system are given by 
\begin{align}
    -i \frac{\partial \psi}{\partial z} =H_{a,b} \psi \ .
\end{align}
The Hamiltonian in momentum space is defined as follows
\begin{align}
H_{a,b} = t\left(
    \begin{array}{cccccc}
    0& \pm 1& 1& 0& \pm e^{-i k}& e^{-i k}\\
    \pm 1& 0& 0& \mp1& 0& 0\\
    1& 0& 0& 1&0& 0\\ 
    0& \mp1& 1& 0& \mp1& 1\\
    \pm e^{i k}& 0& 0& \mp1& 0& 0\\
    e^{i k}& 0& 0& 1& 0& 0
    \end{array}\right) \ ,
\end{align}
where $H_a$ and $H_b$ describe the Hamiltonians for the systems shown in Figs.~\ref{fig1}(a) and (b), respectively. Additionally, $k$ is the horizontal wave-vector or the quasi-momentum. Both Hamiltonians represent the same physical system but reversed, however they do not commute. The propagation constant of each eigenstate is given by its eigenvalue. 
Both Hamiltonians have the same spectrum due to reflection symmetry; Fig.~\ref{fig1}(c) shows the Bloch spectrum which is composed of 6 FBs with eigenvalues $\beta_{\{0,\pm\}}=\{0,\pm 2t\}$, each one with double degeneracy.

The geometrical properties of the FB eigenstates allow the induction of an effective resonator/cavity structure, resulting from the AB caging~\cite{CaceresAravena2022Jun}. The excitation of any lattice site can be decomposed into a superposition of different spatially compact FB modes, implying light oscillation on narrow spatial regions. Figs.~\ref{fig1}(d) and (e) show the different FB modes with an \textit{S} site or a \textit{P} site at the center, respectively, with an amplitude $U$. The eigenstates in Fig.~\ref{fig1}(d) have equal phase for sites in the same column, whereas those in Fig.~\ref{fig1}(e) have a $\pi$ phase difference.

\begin{figure}[t!]\centering
\includegraphics[width=0.47\textwidth]{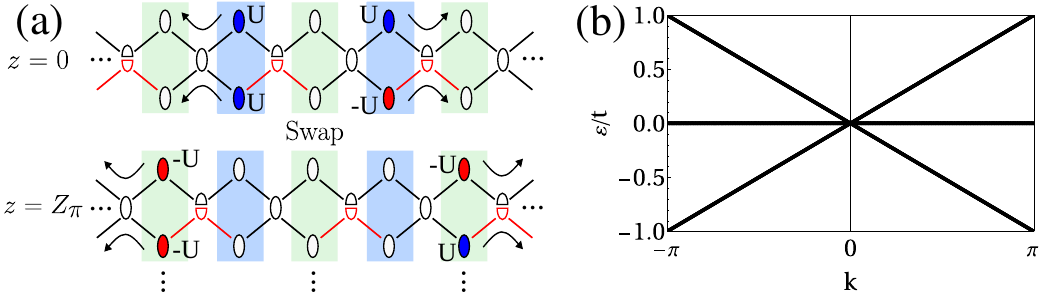}
	\caption{(a) Two initial conditions: in phase (left) and an out of phase (right) profiles at $z=0$, and their evolution after one period $z=Z_\pi=\pi/(2t)$. Blue and red colors represent positive and negative amplitudes, respectively. (b) Floquet quasienergies $\varepsilon$ calculated for a period $Z=2Z_\pi$.}\label{fig2}
\end{figure}

\subsection{Floquet Dynamics}

Fig.\ref{fig2}(a) shows two initial states at $z=0$, one of them in phase (blue zone) and the other one out of phase (green zone). Each of these states is a linear combination of three different FB modes, and each of these states is coupled to the neighboring central waveguides. In order to understand the underlying dynamics, we first analyze how the in-phase (blue-blue) state depicted in Fig.\ref{fig2}(a) propagates to the left. The in-phase state in the blue zone exhibits destructive interference to the right due to the $P$ site; therefore, it does not propagate in that direction. However, it features constructive interference to the left when couple to the $S$ site and, thus, it can translate to the next green zone to the left. Then, the energy transported to the green zone can not continue moving leftwards, because of the destructive interference with the next $P$ site. The evolution of this in-phase initial state in sites can be described using an effective model of three sites (trimer). There, the excitation in one edge transfers entirely to the other edge at half the trimer period, and it returns to the original position after a full period. Consequently, in our protocol, after half a trimer period the central waveguides are swapped (from $S$ to $P$ or from $P$ to $S$) to sustain the propagation. The propagation of the out-of-phase state [see blue-red profile depicted in Fig.\ref{fig2}(a)] follows a similar mechanism, but with an opposite spatial evolution. Due to the initial $\pi$ phase difference, this state constructively interferes at the $P$ site and destructively to the $S$ site. As a consequence, this input condition moves to the right, showing an opposite dynamics compared to the in-phase input condition.

Due to the AB caging effect, the states propagate from the blue zone to the green zone at $z=Z_\pi=\pi/(2t)$. At this point, we swap the central sites to continue the movement of the states, which implies a change in the Hamiltonian from $H_a$ to $H_b$. We leverage the jumping-like effect to design a periodic sequence of the lattices described in Figs.~\ref{fig1}(a) and (b). We denominate them as Masks \textit{a} and \textit{b}, respectively, and switch them by simply exchanging the central waveguides from $S$ to $P$, and vice versa. We repeat this procedure periodically along $z$, considering a total period $Z=2Z_\pi$, as in Fig.~\ref{fig2}(a), with each mask having a length of $Z_\pi$. 

Since the system is composed of a periodic concatenation of masks, the full Hamiltonian is also periodic $H(z+Z)=H(z)$ and, therefore, we analyze it using Floquet theory~\cite{floquet_sur_1883}. Using the static Hamiltonians of both masks, $H_{a}$ and $H_b$, we construct the full $z$-dependent Hamiltonian as
\begin{align}
     H(z) = \left\{ \begin{array}{cc}
        H_a & \text{ if } n Z < z \le (n+1/2) Z \\
        H_b & \text{ if } (n+1/2) Z < z \le \text{$ (n+1)$}Z
    \end{array} \right. \ ,
\end{align}
where $Z$ is the period and $n \in \mathbb{Z}^+$. The treatment of spatially periodic systems is described by Bloch theory~\cite{Bloch1929} (in standard spatially periodic lattices, e.g., crystals) and for dynamically driven systems is described by Floquet theory~\cite{floquet_sur_1883}. The solutions to the $z$-dependent Schr\"odinger equation
\begin{equation}
i \frac{d}{dz} \psi(z) = H(z) \psi(z)
\label{eq:floquet_schroedinger}
\end{equation}
are given by the evolution operator
\begin{equation}
\mathcal{P}(z_2, z_1)
	=\mathcal{T} \exp \left(\frac{1}{i} \int_{z_1}^{z_2} \mathrm{d}z' \ H(z') \right)\ ,
\end{equation}
where $\mathcal{T}$ is the time-ordering operator. 
$\mathcal{P}(z_2, z_1)$ describes the evolution of the system from $z_1$ to $z_2$
\begin{equation}
\psi(z_2) = \mathcal{P}(z_2, z_1) \psi(z_1)\ .
\end{equation}
For a periodic system, the whole temporal evolution can be captured in the stroboscopic evolution operator $\mathcal{P}(Z) = \mathcal{P}(Z, 0)$.
In our case, the period consists of only two distinct phases, so the stroboscopic evolution operator simplifies to
\begin{align}
    \mathcal{P}(Z) = \exp(-i Z H_b / 2) \exp(-i Z H_a / 2)\ .
\end{align}
Now, solving the equation 
$
\mathcal{P}(Z) \psi_\mathrm{F}(z) = \omega_\mathrm{F} \psi_\mathrm{F}(z)
$ 
yields the Floquet eigenstates $\psi_\mathrm{F}(z)$. The Floquet quasienergies $\varepsilon_\mathrm{F}$ are then given by the eigenvalues $\omega_\mathrm{F} = \exp(- i \varepsilon_\mathrm{F} Z)$. The eigenstates are stroboscopically constant [$\psi_\mathrm{F}(z) = \psi_\mathrm{F}(z + Z)$] and provide a useful basis for describing the system dynamics.
Choosing the correct branch cut for the complex logarithm in
\begin{equation}
\varepsilon_n = \frac{i \ln(\omega _n)}{Z}\ ,
\end{equation}
limits $\varepsilon_n$ from $-\pi / Z$ to $\pi / Z$.

Fig.~\ref{fig2}(b) shows the Floquet quasienergies for a Hamiltonian period of $Z=2Z_\pi=\pi/t$, where we observe one flat band at $\varepsilon=0$ and two linear dispersive bands. This feature exhibits a Dirac cone-like dispersion. The Floquet band spectrum is periodic in energy, so the dispersive bands go linearly from $\varepsilon=0$ to $\pm 2t$ in the range $k=0$ to $2\pi$. 
The excitation of a Floquet band with constant slope allows a wave packet to propagate without dispersion ($\partial^2 \varepsilon/\partial k^2=0$), while maintaining its compact shape. Moreover, an excitation belonging to one of these dispersive bands will translate through the lattice at a constant velocity until reaching the end of the lattice. 
The Floquet dispersionless traveling states are composed of linear combinations of FB modes from the static system. These states have a vertical geometry and four possible input configurations per unit cell. Linear combinations with complex phase of these four inputs are also possible. To create a traveling excitation in one of the dispersive bands, two adjacent vertical waveguides must be excited simultaneously with equal amplitude, either in-phase or out-of-phase. In Figs.~\ref{fig1}(a) and (b) we highlight in blue and green the vertical waveguides that host Floquet dispersionless states. The direction of movement depends on the input phase and input location. An in-phase excitation of the vertical waveguides highlighted in blue in Fig.~\ref{fig1}(a) will start moving to the left. After periodically swapping the masks the localized state will keep moving in the same direction. On the other hand, if the same waveguides are excited now in an out-of-phase configuration, the wave packet will translate to the right and will continue moving in that direction. If we now choose the other unit cell region [green area in Fig.~\ref{fig1}(a)], a reversed dynamics will be observed. Given that the in-phase and out-of-phase excitations are compact photonic states of the Floquet system, these states will move independently from each other.

\subsection{Topological Classification}
%
In a one-dimensional (1D) system the topological features are characterized by their topological invariant, edge states, and symmetries~\cite{Altland_1997}. 
For a Floquet driven system the topological classification is performed according to their Floquet evolution operator. We start by studying the symmetries of the system. 
The symmetry in the bands suggest that there might be a chiral symmetry, thus we should find a chiral operator $\mathcal{C}$ that satisfy $\mathcal{C}\mathcal{P}(Z)\mathcal{C}^\dagger=\mathcal{P}^\dagger(Z)$. 
In order to find the chiral operator $\mathcal{C}$ of the evolution operator, we first calculate the chiral operator $\Gamma$ of the static Hamiltonians as:
\begin{align}
    \Gamma = \left(
    \begin{array}{cccccc}
        -1 & 0 & 0 & 0 & 0 & 0\\
        0 & 1 & 0 & 0 & 0 & 0\\
        0 & 0 & 1 & 0 & 0 & 0\\
        0 & 0 & 0 & -1 & 0 & 0\\
        0 & 0 & 0 & 0 & 1 & 0\\
        0 & 0 & 0 & 0 & 0 & 1
    \end{array}
    \right)\ .
\end{align}
Note that $\Gamma=\Gamma^\dagger=\Gamma^{-1}$. 
We check that this is the chiral operator of both static Hamiltonians, this is: $\Gamma H_{a,b}=-H_{a,b} \Gamma$. Now, we need to calculate 
$
 \Gamma\mathcal{P}(Z)= \Gamma e^{-i Z H_b / 2} e^{-i Z H_a / 2} 
$, so we proceed to open the exponential as follows
\begin{align}
      \Gamma e^{-i Z H_b / 2}  &=
     \sum_{n=0}^\infty \frac{(-i Z/2)^n}{n!} \Gamma(H_b)^n \nonumber \\
     &=\sum_{n=0}^\infty \frac{(-i Z/2)^n}{n!} (-H_b)^n \Gamma
     = e^{iZH_b/2}\Gamma\ .
\end{align}
In this manner, $\Gamma\mathcal{P}(Z) =  e^{i Z H_b / 2} e^{i Z H_a / 2} \Gamma$.

 Now, we find the exchange operator $\mathcal{S}$ that exchanges the Hamiltonian $a$ for the $b$, and vice versa; this is: $\mathcal{S}H_{a,b}\mathcal{S}^\dagger=H_{b,a}$. The exchange operator is given by:
\begin{align}
    \mathcal{S} = \left(\begin{matrix}-1 & 0 & 0 & 0 & 0 & 0\\0 & 1 & 0 & 0 & 0 & 0\\0 & 0 & -1 & 0 & 0 & 0\\0 & 0 & 0 & -1 & 0 & 0\\0 & 0 & 0 & 0 & 1 & 0\\0 & 0 & 0 & 0 & 0 & -1\end{matrix}\right)\ .
\end{align}
We find that the chiral operator of the Floquet system is $\mathcal{C}=\mathcal{S}\Gamma$ and
\begin{align}
    \mathcal{C}\mathcal{P}(Z)\mathcal{C}^\dagger &= \mathcal{S}\Gamma\mathcal{P}(Z) \mathcal{C}^\dagger \\ 
    &=  e^{i Z H_a / 2} e^{i Z H_b / 2} \mathcal{S}\Gamma\mathcal{C}^\dagger \\ 
    &= \mathcal{P}^\dagger(Z) \ .
\end{align}
Therefore, our Floquet driven system possesses chiral symmetry. After a little bit of algebra, we have found an interesting physical meaning behind this symmetry: the in-phase and the out-of-phase states, which travel in opposite directions, are chiral pairs (one is the chiral pair of the other).

The system also has time-reversal symmetry which is given by
\begin{align}
    \Theta \mathcal{P}(Z,-k) \Theta^\dagger =  \mathcal{P}^\dagger(Z,k) ,
\end{align}
and the particle-hole symmetry which is given by
\begin{align}
    \Lambda \mathcal{P}(Z,-k) \Lambda^\dagger =  \mathcal{P}(Z,k) .
\end{align}

The time-reversal symmetry operator in this system is $\Theta=\mathcal{S}K$, where $K$ is the complex conjugate operator. 
We check that
\begin{align}
    \mathcal{S}K\mathcal{P}(Z,-k)K^\dagger \mathcal{S}^\dagger &= \mathcal{S} (e^{-i Z H_b(-k) / 2} e^{-i Z H_a(-k) / 2})^*\mathcal{S}^\dagger\\
    &= e^{i Z H_a(k) / 2} e^{i Z H_b(k) / 2}\\
    &= \mathcal{P}^\dagger(Z,k) \ ,
\end{align}
and we check that $\Theta\Theta^\dagger=1$.

The particle-hole symmetry operator is $\Lambda=\Gamma K$, and when we apply the symmetry operator:
\begin{align}
    \Gamma K\mathcal{P}(Z,-k)K^\dagger \Gamma^\dagger &= \Gamma  (e^{-i Z H_b(-k) / 2} e^{-i Z H_a(-k) / 2})^*\Gamma^\dagger\\
    &= \Gamma  e^{i Z H_b(k) / 2} e^{i Z H_a(k) / 2}\Gamma^\dagger\\
    &= e^{-i Z H_b(k) / 2} e^{-i Z H_a(k) / 2}\\
    &= \mathcal{P}(Z,k) \ ,
\end{align}
and we check that $\Lambda\Lambda^\dagger=1$. 

This set of symmetries suggests that the system might belong to the BDI class of topological invariants. In order to correctly classify it, we must either find the edge states and the topological invariant which in this case is the winding number. 
The bulk-boundary correspondence guarantees the existence of edge states when the topological invariant is non-trivial, and for the case of Floquet driven systems this relation also holds true~\cite{PhysRevB.82.235114}. The static Hamiltonians $H_{a,b}$ have edge states; however, for the Floquet topological classification we have to find the edge states from the Floquet evolution operator. 
By diagonalizing the Floquet evolution operator for an open chain, we find two interesting states at quasienergy $\varepsilon=0$. The spatial intensity of these states peaks at the edges and decays to the bulk, but they lack the exponential decay that topological edge states exhibit. This kind of decay resembles the decay observed in topological edge states when they delocalize as the gap starts to close and the system is approaching a critical point. This led us to think that we might be in a critical point so we aim to open a gap. Therefore, we introduce a dimerization to the lattice of a Su–Schrieffer–Heeger-type, and we observe that these two states become full edge states when a gap opens at energy $\varepsilon=0$. Even with dimerization these states are at quasienergy $\varepsilon=0$. 

For a BDI Floquet 1D topological insulator, the topology is characterized by two winding numbers~\cite{PhysRevB.88.121406,koyama_designing_2023}: $W_0$ which counts zero-quasienergy edge states and $W_\pi$ which counts edge states of quasienergy $\varepsilon=\pi/Z$. To perform this calculation in a Floquet system composed of 2 static Hamiltonians, we must shift the time origin to the midpoint of each driving step to construct two time-symmetric evolution operators:
\begin{align}
    \mathcal{P}_1(k)&=e^{-iZH_a(k)/4}e^{-iZH_b(k)/2}e^{-iZH_a(k)/4}\ ,\\
    \mathcal{P}_2(k)&=e^{-iZH_b(k)/4}e^{-iZH_a(k)/2}e^{-iZH_b(k)/4}\ ,
\end{align}
these operators have the form:
\begin{align}
    \mathcal{P}_j(k) = \left( 
    \begin{array}{cc}
       A_j(k)  & B_j(k)  \\
       B_j^\dagger(k)  & A_j(k)
    \end{array}
    \right)\ ,
\end{align}
and now we compute
\begin{align}
    W_j = \frac{1}{2\pi i }\int_{-\pi}^{\pi} dk \; \frac{d}{dk}\ln(\det(A_j(k)+iB_j(k)))\ ,
\end{align}
and finally we obtain $W_0=(W_1+W_2)/2$ and $W_\pi=(W_1-W_2)/2$. This calculation for our system leads to $W_0=-0.5$ and $W_\pi=0$. Usually this calculation leads to an integer number and it's related to the amount of edge states of the system, however as discussed previously the system is in a critical point and the topological invariant is ill-defined in this point. 
This is because the topological edge states are related to the opening of bands, and in our system we observe Dirac-like cones with crossing bands.

The topological invariant of a periodically driven 1D system is the winding number $\nu$. 
We calculate the winding number $\nu_n$ corresponding to the $n$th Floquet quasienergy $\varepsilon_n$ using the expression in~\cite{PhysRevB.82.235114}:
\begin{align}
    \nu_n = \frac{1}{2\pi} \int_{-\pi}^\pi dk \frac{d\varepsilon_{n}(k)}{dk} Z \ ,
\end{align}
where $Z=\pi/t$ is the period, and the total winding number is given by $\nu=\sum_n \nu_n$.  
We calculate $\nu$ for each band and obtain
\begin{align}
    \nu = \left\{ 
    \begin{array}{cc}
        0  & \text{for $\varepsilon=0$,  }\\
        1  & \text{for $\varepsilon=k t/\pi$, } \\
        -1 & \text{for $\varepsilon=-k t/\pi$, }
    \end{array}
    \right. 
\end{align}
all with a double degeneracy. The total winding number of the system is zero. However, because of the band crossing the topological invariant is ill-defined and this result does not constitute a globally defined band-resolved winding number in the usual sense.

\section{Experimental Floquet mobility}

\begin{figure}
    \centering
    \includegraphics[width=0.45\textwidth]{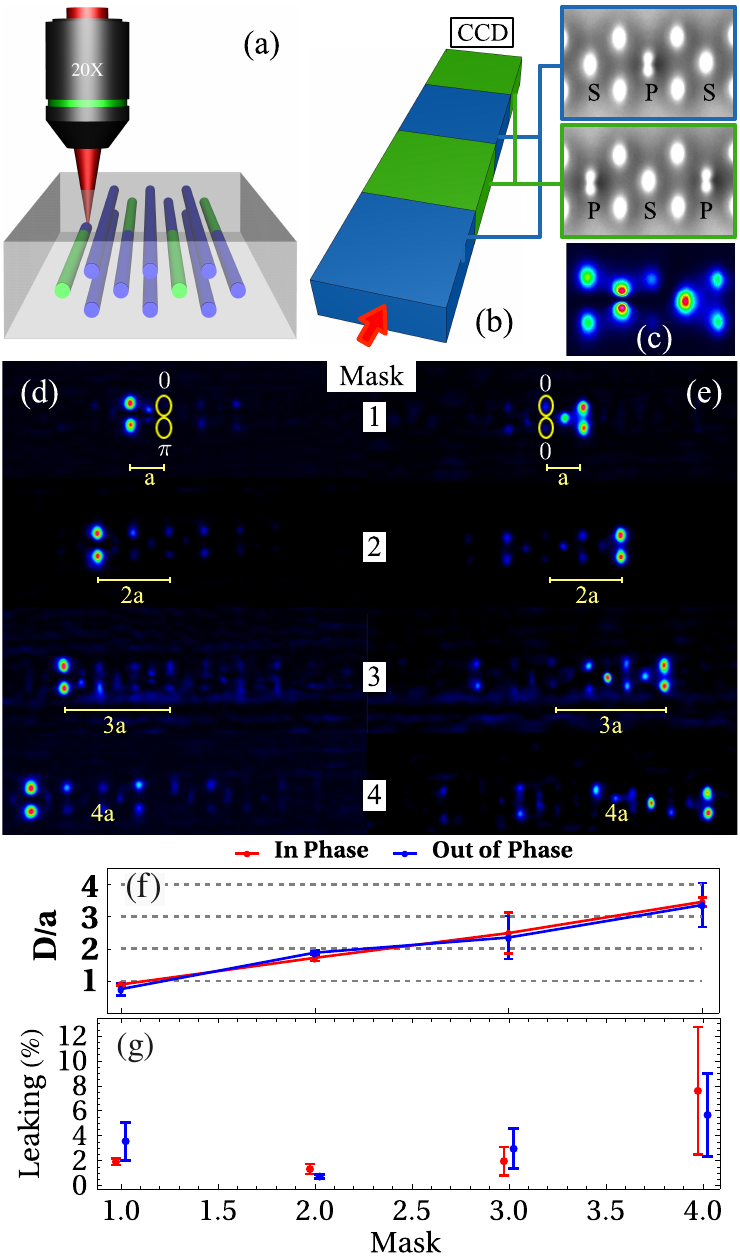}
    \caption{(a) Sketch of the femtosecond laser writing technique alternating the central waveguides between $S$ (blue) and $P$ (green) types. (b) The alternation of the central waveguides defines two different mask segments highlighted in blue and green. Insets show a bright-field microscopy image of each mask. (c) Output intensity image showing the $S$ and $P$ modes. (d)-(e) Intensity output at the end of each mask segment $\{1,2,3,4\}$, for the (d) out-of-phase initial condition and the (e) in-phase condition. (f) Normalized displacement $D/a$ of the wave packet from its initial condition after each mask segment.(g) Percentage of leaking per mask. For (f) and (g) red (blue) dots are for the in-phase (out-of-phase) condition.}
    \label{fig:experimental_fig}
\end{figure}

We use the femtosecond (fs) laser writing technique~\cite{Szameit2010} to fabricate the SP diamond photonic lattices under study on a borosilicate glass wafer of length $L=70$ mm. By adjusting the writing power, we control the refraction index contrast and fabricate $S$-only as well as multimode \textit{P} waveguides. Specifically, the sample is translated by a nanometer resolution XYZ Aerotech stage at a fixed writing velocity of $0.8$ mm/s, with a nominal writing power of $15.31$ and $17.50$ mW for the $S$ and $P$ waveguides, respectively. A fine tuning protocol for the writing power is implemented such that a second order $P$ photonic mode effectively interacts with a fundamental $S$ mode~\cite{guzman-silva_experimental_2021}. For this, we match the propagation constants of the \textit{S} and the $P$ states at different lattice positions, for the specific excitation wavelength of $640$ nm. Fig.~\ref{fig:experimental_fig}(a) shows a sketch of the fabrication technique, where the blue (green) waveguides correspond to the $S$ ($P$) sites.

Figures \ref{fig1}(a) and (b) show that only the central $A$ and $D$ sites have to be exchanged for the two considered masks, while the sites $B$, $C$, $E$, and $F$ are always kept as $S$ sites during the fabrication [see sketch in Fig.~\ref{fig:experimental_fig}(a)]. After a distance of $Z_{\pi}=16.1$ mm (obtained after a full optimization of the lattice geometry and coupling constants), the central waveguides are swapped and we create the two required lattice configurations [see the microscope images in Fig.~\ref{fig:experimental_fig}(b)]. The implementation of this protocol, that exchanges $H_a$ and $H_b$, produces an abrupt change in the core refractive index of the central waveguides. This can induce losses and imperfect mode conversion if and only if there is a relevant amount of light at the central waveguides during the swap. As the swap is inserted at $z=Z_{\pi}$ in our fabricated lattices, the light has already passed though this region and the protocol is successfully implemented. 

By using an image setup~\cite{grafeno24} (based on spatial light modulators), we modulate in amplitude and phase a broad laser beam at a wavelength of $640$ nm. In this way, we create the nontrivial in-phase and out-of-phase input conditions, described above. The initial condition is injected at the input facet of the glass wafer, as indicated by a red arrow in Fig.~\ref{fig:experimental_fig}(b). Then, the light travels along $z$, and we image the output profiles at the glass output facet using a $10\times$ microscope objective and a beam profiler. Fig.~\ref{fig:experimental_fig}(c) shows an experimental image, in which we clearly observe the $S$ and the $P$ photonic modes.

We study the dynamics along $z$ by implementing a $z$-scan method~\cite{guzman-silva_experimental_2021,CaceresAravena2022Jun,grafeno24}. For this, the input waveguides are fabricated along the whole glass wafer (length $L$), while the rest of the lattice is fabricated with a shorter propagation length. Therefore, once the input excitation sees the lattice it starts to interact with the system. As our half period was set to $Z_{\pi}=16.1$ mm, in a glass wafer of length $L=70$ mm, we could insert up to four concatenated masks, as it is sketched in Fig.~\ref{fig:experimental_fig}(b). Figure ~\ref{fig:experimental_fig}(d) shows intensity output profiles obtained after one, two, three, and four concatenated masks for an out-of-phase input excitation (yellow circles). In this case, we observe a clear transfer of a compact localized excitation, which has been translated to the left. This observation is indeed quite relevant in physics and photonics, as it corresponds to a controlled transport of a compact localized state along a discrete lattice. This observation is demonstrated in a purely linear system under periodical driving, something that was only suggested in Ref.~\cite{CaceresAravena2022Jun} and experimentally demonstrated in the present work. For several years, the use of nonlinear phenomena was thought to be the only solution for observing compact transport in discrete systems~\cite{PNPKivshar,Vicencio01,LEDERER20081}. However, our observation shows that multiorbital coupling and Floquet engineering could be combined to achieve discrete and controlled translation of compact wavepackets on linear lattices. Fig.~\ref{fig:experimental_fig}(e) shows a similar evolution for an in-phase initial condition (yellow circles), where we observe that the light is effectively translated to the right. 

We determine the averaged discrete translation of the wavepacket by estimating its displacement $D$, defined as:
\begin{align}
D = \frac{1}{N} \sum_{i}^N |X_i -n_i|\ ,
\end{align}
where $N=6$ is the number of experiments for the same initial condition, but shifted across the lattice. $X_i$ is the wave packet center of mass and $n_i$ is the position of the initial condition, for any $i$-th experimental realization. Figure~\ref{fig:experimental_fig}(f) shows the absolute value of the normalized field displacement $D$ in units of the distance $a$ (horizontal distance among vertical waveguides), obtained after applying a given mask. The error bars describe the standard deviation for these measurements. We observe very good discrete mobility with clear values around $1,\ 2,\ 3$, and $4$, validating the implemented Floquet protocol.

To quantify the deviation from the ideal expected evolution, we define the leaking fraction $L$ as
\begin{align}
    L=100\%\frac{(\text{PR}_\text{exp}-\text{PR}_\text{theo})}{(N_\text{total}-\text{PR}_\text{theo})}\ ,
\end{align} 
where $N_\text{total}$ is the total number of sites, while $\text{PR}_\text{exp}$ and $\text{PR}_\text{theo}$ represent the experimental and theoretical participation ratio, respectively.
The participation ratio is defined as 
\begin{align}
    \text{PR}=\frac{\left(\sum_n |u_n|^2\right)^2}{\sum_n |u_n|^4}\ . 
\end{align}
For three or less masks we obtain very good results with an average leaking of around $3\%$. We observe that the leaking increases when adding more masks, reaching around $7\%$ for both conditions after four masks. Also, we notice that the error bar becomes significantly larger, as a manifestation of accumulative fabrication defects and incomplete transfers occurring after several caging periods.

\section{Conclusion}

In conclusion, we have investigated a Floquet scheme on an all-FB lattice. Without periodically driving the Hamiltonian, we would observe only caging effects due to an all-FB spectra~\cite{CaceresAravena2022Jun}. However, by engineering the lattice along the propagation coordinate, we transformed the dynamical properties and observed the propagation of highly compact localized wavefunctions. 
From the obtained topological features, we demonstrated that the traveling modes are chiral states.  The static Hamiltonians exhibit chiral symmetry and share the same chiral operator. The Floquet driven system is also chiral and its corresponding chiral operator is related to the chiral operator of the static systems. We also find interesting states at $\varepsilon= 0$ that become edge states when the gap is opened, so out system is in a transition point. 
We experimentally implemented the Floquet protocol in femtosecond laser written photonic lattices. We steered compact localized states through the system using four masks, and observed a controlled and discrete transport of the wavepackets. Our results show that FB localized states can be used to propagate energy across a given lattice, where the direction of transport can be completely determined by selecting the input phase and/or input position. Recent experimental work has explored the transport of discrete wavepackets~\cite{refId0}, specifically by bending the waveguides, but producing also extra losses~\cite{SnyderLove:1983,Marcuse:76,Dai:04}. In our work, we extend these frontiers by only using evanescent coupling without the need of curving the waveguides. We exchange the central sites, that are used to create the interaction between the vertical waveguides, generating compact transport of flat band zero tail modes. We calculated the leaking fraction of our implementation and we find that up to three masks we observe a very good average leakage lower than $3\%$. However, we observe an increasing leaking fraction of around $7\%$ when adding the 4th mask. This can be explained from the increment of the propagation distance and the introduction of accumulative errors for the completed light jumps, including small unbalanced couplings. For larger propagation distances the next-next-nearest-neighbor interactions could also start to play a role, affecting the chiral lattice symmetry and the expected compact transport of energy.

\begin{acknowledgments}
This research and publication is funded by the Deutsche Forschungsgemeinschaft (DFG, German Research Foundation) through IRTG 2676/1 ‘Imaging of Quantum Systems’, project no. 437567992, the Millennium Science Initiative Program ICN17\_012 and FONDECYT Grant No. 1231313. The authors thanks to Dr. Rafael Molina for helpful discussions.
\end{acknowledgments}

\appendix

\bibliography{bibliography}

\end{document}